\documentclass[12pt]{article}
\usepackage{amsmath}
\usepackage{graphicx}
\usepackage{enumerate}
\usepackage[nofiglist,notablist]{endfloat}
\usepackage{url} 
\usepackage{authblk}
\usepackage{multirow}
\usepackage{booktabs}
\usepackage[T1]{fontenc}
\usepackage[colorlinks, linkcolor = blue, citecolor = blue]{hyperref}
\usepackage{newtxmath,newtxtext}
\usepackage{color}
\usepackage{url}
\usepackage[normalem]{ulem}

\usepackage{achemso}
\setkeys{acs}{articletitle = true}

\begin{document}

\def\spacingset#1{\renewcommand{\baselinestretch}%
{#1}\small\normalsize} \spacingset{1}
\setlength{\bibsep}{0pt plus 0.8ex}


\title{\bf Sample size re-estimation in Phase 2 Dose-Finding: Conditional power vs. Bayesian predictive power}
\author[1]{Qingyang Liu}
\author[2]{Guanyu Hu\thanks{To whom correspondence should be addressed: guanyu.hu@missouri.edu}}
\author[3]{Binqi Ye}
\author[4]{Susan Wang}
\author[5]{Yaoshi Wu}
\affil[1,5]{University of Connecticut, Storrs, CT, USA}
\affil[2]{University of Missouri - Columbia, Columbia, MO, USA}
\affil[3]{Boehringer Ingelheim (China), Shanghai, China}
\affil[4]{Boehringer-Ingelheim Pharmaceutical Inc., Ridgefield, CT, USA}
\date{}
\maketitle
\bigskip

\begin{abstract}
Unblinded sample size re-estimation (SSR) is often planned in a clinical trial when there is large uncertainty about the true treatment effect. For Proof-of Concept (PoC) in a Phase II dose finding study, contrast test can be adopted to leverage information from all treatment groups. In this article, we propose two-stage SSR designs using frequentist conditional power (CP) and Bayesian predictive power (PP) for both single and multiple contrast tests. The Bayesian SSR can be implemented under a wide range of prior settings to incorporate different prior knowledge. Taking the adaptivity into account, all type I errors of final analysis in this paper are rigorously protected. Simulation studies are carried out to demonstrate the advantages of unblinded SSR in multi-arm trials. 
\end{abstract}

\noindent%
\vfill

\newpage
\spacingset{1.5} 

\section{Introduction}\label{sec:intro}
Sample size calculation is a key step in the planning of a clinical trial. The sample size needs to be large enough to achieve the objective of the clinical trial, and at the same time should not be too big for ethical, clinical or economical reasons. The sample size of a trial depends on the study design, the hypothesized treatment effect, the variability of the study endpoint, the type I error rate, and the power. Very often, especially in some new disease areas and during the very early stage of a trial, the most critical parameter for sample size calculation, the treatment effect of the study endpoint is not precisely known. The uncertainty may result in an under- or overpowered trial, leading to potential risk of trial failure or costly large trial. This risk can be alleviated by sample size re-estimation (SSR), a popular adaptive approach to correct for incorrectly specified sample sizes based on updated knowledge. Rather than an one-stage fixed study design, SSR re-assesses the sample size at interim analysi(e)s in the mid-course of the trial. This idea of adaptive sample size increase allows us to start with a provisional sample size calculation. After recruiting a portion of the planned sample, we can choose to increase the sample size provided the interim treatment effect estimate is a little worse than expected. 

SSR can be done blinded or unblinded. With the blinded SSR, only nuisance parameters are re-estimated to update the sample size calculation. Such nuisance parameters include common variance of continuous endpoints \citep{shih_1998} and pooled event rate of binary endpoints \citep{gould_1992}, etc. It has been shown that blinded SSR procedure does not impact the type I error rate for most superiority trials \citep{kieser_2003}, but may lead to type 1 error inflation in the setting of non-inferiority or equivalence studies \citep{friede_2003}. According to a recent survey by \citet{morgan_2014}, unblinded SSR is more popular in confirmatory trials from industry compared with blinded SSR. As for unblinded SSR, the treatment effect assumption can be updated during the interim analysis, but at the same time the type I error must be carefully controlled to maintain the statistical rigor of trial results \citep{mehta_2006,proschan_2009}. There have been a variety of methods introduced to protect type I error for adaptive group sequential designs, such as combination of p-values \citep{lehmacher_1999}, weighted test statistics \citep{cui_1999}, and the conditional error approach \citep{muller_2004}. Usually, the decision rule on increasing sample size is based on the probability of success given the current trend of interim data, the conditional power \citep{mehta_2011} from the frequentist perspective, or the Bayesian predictive probability of study success \citep{wang_2007, zhong_2013}. Compared with the frequentist definition of power, the Bayesian power approach utilizes prior information flexibly and provides uncertainty modeling of the unknown parameters. \citet{chuang_2006} and \citet{pritchett_2015} discussed comprehensively on both blinded and unblinded SSR in methodological, regulatory and practical aspects. 

Recently, more and more Phase II designs conduct Proof-of-Concept (PoC)  to make ``go/no-go'' decision and dose ranging to characterize the dose-response (D-R) relationship in one study to speed up the clinical development process \citep{bretz_2005,wang_2012}. To detect efficacy signal from multiple doses, one possible method is to do pairwise comparisons with placebo. As suggested by the ICH guideline \citep{ich_1994}, another possibility is to leverage information from all treatments to perform a single contrast test to show trend in a specific direction \citep{ruberg_1995}. Power of contrast tests depends on the true underlying D-R relationship. Getting this wrong may result in a loss in power. As a consequence, more robust tests based on the maximum of multiple contrasts have been proposed to reduce the uncertainty of power under different D-R profiles \citep{bretz_2005}. 

In this paper, we are interested in applying unblinded SSR to multiple contrast test. Our approach is on the basis of a general two-stage adaptive dose-finding study design with multiple contrast test proposed by \citet{miller_2010}. With rigorous familywise type I error rate control, this design framework incorporates data dependent adaptions like sample size increase, alterations of allocation proportion and contrasts. Nonetheless, we focus our study on SSR only in this article. Using conditional power (CP) and Bayesian predictive power (PP) to assess interim data, our sample size re-estimation rules are motivated by the key idea of ``promising zone''. We use simulation studies to compare the practical performance of frequentist and Bayesian powers in our SSR design. 

The rest of this paper is organized as follows. In Section~\ref{sec:frequentist}, we present the frequentist SSR design for single contrast test first, and then generalize our design to multiple contrast test. In addition, the procedure for type I error protection is also discussed in Section~\ref{sec:frequentist}. Next, parallel Bayesian SSR with detailed discussions on Bayesian computation techniques and multiple testing extension is proposed in Section~\ref{sec:bayes}. In Section~\ref{sec:simu}, simulation studies are carried out to validate the performance of our frequentist and Bayesian designs. At last, we conclude this paper with a brief discussion in Section~\ref{sec:discuss}.

\section{Frequentist approach for SSR}\label{sec:frequentist}
\subsection{Contrast test}
For a dose-finding study with continuous efficacy endpoint, suppose the total number of randomized treatment groups is $k$ ($k\geq 3$) and the total sample size is $N$. Let $Y_{ij}$ denote the response of the $j$th patient from the $i$th treatment group. The sample allocation ratio $\boldsymbol{\phi}=(\phi_1,\ldots,\phi_k)^\top$ is a $k$-dimensional sum-to-one vector. The analysis of variance (ANOVA) model has the following form: 
\begin{equation*}
Y_{ij}=\mu_i+\epsilon_{ij}, \; \text{ where } \epsilon_{ij}\overset{\text{i.i.d.}}{\sim} \mathcal{N}(0,\sigma^2),\; i=1,\ldots,k, \text{ and } j=1,\ldots,N\phi_i.
\end{equation*}
We assume the common variance $\sigma^2$ to be a known constant for the normally distributed data. Denote the mean vector $(\mu_1,\ldots,\mu_k)^\top$ by $\boldsymbol{\mu}$, and the collection of all observed response by $\boldsymbol{Y}$.

Linear contrast test, also known as trend test, is a common choice to leverage clinical outcomes from all arms so as to detect signal of non-flat dose-response relationship. To test against the null hypothesis $H_0: \sum_{i=1}^k c_i\mu_i=\boldsymbol{c}^\top\boldsymbol{\mu}=0$, the conventional trend test statistic is: 
\begin{equation*}
T=\frac{\sum_{i=1}^k c_i\overline{Y}_{i.}}{\sigma\sqrt{\frac{1}{N}\sum_{i=1}^k  c_i^2/\phi_i}}\sim\mathcal{N}\left(\frac{\sum_{i=1}^k c_i\mu_i}{\sigma\sqrt{\frac{1}{N}\sum_{i=1}^k  c_i^2/\phi_i}}\;,\; 1\right), 
\end{equation*}
where $\boldsymbol{c}$ is the $k$-dimensional contrast vector satisfying $\sum_{i=1}^k c_i=0$. We use $\widehat{\delta}=\sum_{i=1}^k c_i\overline{Y}_{i.}$ to denote the point estimate of leveraged treatment effect. Without any loss of generality, suppose large values of leveraged treatment effect indicates desired drug effect. Then, for one-sided trend test, the null hypothesis is rejected if $T>\Phi^{-1}(1-\alpha)=z_\alpha$, where $\Phi^{-1}$ is the inverse cumulative distribution function of standard normal distribution and $\alpha$ is the significance level. 

The contrast vector can be optimized to maximize the power. In order to specify $\boldsymbol{c}$, a possible strategy is assuming a likely shape of the mean vector, namely $\boldsymbol{\mu}^0$, and then adopting the most powerful linear contrast, of which the derivation has been discussed by the work of \citet{bretz_2005} as well as \citet{pinheiro_2014}. Also depending on the allocation ratio, the optimal contrast $\boldsymbol{c}$ can be computed through
\begin{equation*}
c_i\propto \phi_i(\mu^0_i-\overline{\mu}) ,\quad i=1,\ldots,k, 
\end{equation*}
where $\overline{\mu}=\sum_{i=1}^k \phi_i\mu^0_i$, and the contrast can be uniquely determined by imposing the constraint $\|\boldsymbol{c}\|_2=1$. Information like literature data of related compounds, pre-clinical data, pharmacokinetic or pharmacodynamic data from Phase I studies can be used to determine the contrast vector.

\subsection{Type I error control of two-stage adaptive design}
Consider the fixed design of a two-stage study with $N_1$ and $N_2$ observations in stage 1 and 2, respectively, and a common sample allocation ratio $\boldsymbol{\phi}$ for both stages, and with no adaptions. A contrast test statistic can be computed for each stage separately only with data from that stage. Under the same contrast vector for both stages, the trend test statistic for stage $j$ is denoted by $T_j$, $j=1,2$. At final analysis, the trend test statistic is equivalent to the following combination of two stages,
\begin{equation*}
T^*=r^{1/2}\cdot T_1+(1-r)^{1/2}\cdot T_2, 
\end{equation*}
with $r=N_1/(N_1+N_2)$ \citep{miller_2010} to be the information fraction. Please note that in order to protect the type I error, $N_1$, $N_2$, $\boldsymbol{\phi}$, $\boldsymbol{c}$ should be pre-determined and there should be no adaptive modifications on the stage 2 setting when breaking blind at the interim analysis. 

Now suppose that the SSR at interim changes the sample size in stage 2 from $N_2$ to $\widetilde{N_2}$ on the basis of stage 1 result $\boldsymbol{Y}^{(1)}$, while the contrast vector and allocation ratio remain fixed, so that the refined stage 2 trend test statistic turns into
\begin{equation*}
\widetilde{T}_2=\frac{\sum_{i=1}^k c_i\overline{Y}_{i.}^{(2)}}{\sigma\sqrt{\frac{1}{\widetilde{N}_2}\sum_{i=1}^k  c_i^2/\phi_i}}. 
\end{equation*}
To prevent the type I error inflation caused by data dependent sample size adaption, the combination test statistic should be weighted by the pre-declared sample size design
\begin{equation}
\widetilde{T}^*=r^{1/2}\cdot T_1+(1-r)^{1/2}\cdot \widetilde{T}_2, 
\label{reweighted}
\end{equation}
and positivity should be declared if $\widetilde{T}^*>z_\alpha$. In additional to adaptive sample size increase, this combination statistic is also applicable to changes in second stage contrast vector and allocation ratio with rigorous type I error control.

\subsection{Sample size re-estimation} \label{sec:ssr}
With pre-specified $\alpha$-level, statistical power, and population variance, the sample size calculation depends on the estimation of effect size, which is unknown prior to the trial. Even with much information from pre-clinical and Phase I data, there is still substantial uncertainty on the accuracy of treatment effect estimation. Therefore, to prevent unexpected power loss, it is often beneficial to re-assess the effect size and update the planned sample size through an interim analysis. Unblinded SSR provides an opportunity for the study to maintain adequate power to detect efficacy, even if the initial treatment effect estimate is too optimistic. In the following part, we introduce an adaptive design for SSR based on CP. 

At the design stage, appropriate planning is required for unblinded SSR by the FDA guidance on adaptive designs \cite{fda_2019}. In our setting, the planned interim analysis is conducted when $N_1$ subjects have been enrolled and evaluated. The sample allocation proportion is $\boldsymbol{\phi}$ at that time. The original residual sample size is $N_2$ with the same allocation as stage 1. By assuming an optimistic scenario red in terms of treatment effect, this planning can be easily fulfilled by traditional sample size calculation. At the interim look, we are able to decide whether to increase the stage 2 sample size or stick to the initial plan. 

When interim data is available, the uncertainty is now restricted to the result from the second stage. CP is defined as the conditional probability of trial success provided the stage 1 result. Note that CP also depends on the treatment effect size with respect to contrast $\boldsymbol{c}$: $\delta=\sum_{i=1}^k c_i\mu_i$. It takes both the existing evidence at the time of interim analysis and the uncertainty of future data into consideration. Based on the weighted combination statistic \eqref{reweighted}, CP is determined by the following formula \citep{lan_2009,wassmer_2016}: 
\begin{equation}
\text{CP}(n_2,\delta\mid\boldsymbol{Y}^{(1)})=1-\Phi\left(z_\alpha(1-r)^{-1/2}-\left(\frac{r}{1-r}\right)^{1/2}T_1-\frac{\delta\sqrt{n_2}}{\sigma\sqrt{\sum_{i=1}^k c_i^2/\phi_i}}\right), 
\label{cp}
\end{equation}
where $n_2$ is the sample size of stage 2 treated as a variable. It is obvious to find that the CP function is an increasing function of $n_2$ approaching probability 1 provided that $\delta$ is positive. When $n_2$ is fixed, the function is increasing in $\delta$. To form the decision rule for sample size increase, some necessary assumption on the value of $\delta$ is in need. It could be the one calculated from the originally anticipated responses, or the effect estimated from the current data $\widehat{\delta}=\sum_{i=1}^k c_i \overline{Y}_{i.}^{(1)}$ \citep{lachin_2005}. 

Based on a partition according to CP, \citet{mehta_2011} proposed a decision rule for SSR, in which the sample size adjustment occurs when the result at the interim analysis is promising. Practically, due to consideration of budget and enrollment feasibility, also to prevent the test from being too sensitive so that clinically meaningless signal can be easily detected, the sample size for stage 2 cannot be unlimitedly large. Therefore, we let $N_{\text{max}}$ be a pre-specified maximum number of patients allowable for stage 2. It can be calculated from the worst D-R scenario, for which investment would still be considered worthwhile. Using the reweighted final statistic, a similar decision rule described by three CP zones is introduced below.
\begin{enumerate}
\item Unfavorable: $\text{CP}(N_2)<\text{CP}_{\text{min}}$ or $\delta<0$. The stage 2 sample size is maintained to be $N_2$. 
\item Promising: $\text{CP}_{\text{min}}\leq\text{CP}(N_2)<1-\beta$. The CP in \eqref{cp} is used to determine the stage 2 sample size. By solving $\text{CP}(n_2)=1-\beta$, we are able to derive the new sample size: 
\begin{equation*}
\widetilde{N}_2=\min\left\{\left(\sum_{i=1}^k  c_i^2/\phi_i\right)\cdot\frac{\sigma^2}{\delta^2}\cdot\left[z_\beta+z_\alpha(1-r)^{-1/2}-\left(\frac{r}{1-r}\right)^{1/2}T_1\right]^2,N_\text{max}\right\}. 
\end{equation*}
\item Favorable: $\text{CP}(N_2)\geq 1-\beta$. The stage 2 sample size is maintained to be $N_2$. 
\end{enumerate}

The lower bound of the ``promising zone'', $\text{CP}_{\text{min}}$, is typically chosen from 30\% to 50\% \citep{mehta_2011}. Increasing $\text{CP}_{\text{min}}$ narrows the ``promising zone'', so that the overall power gain may be too small. On the other hand, reducing the value of $\text{CP}_{\text{min}}$ may result in higher chances of unnecessary sample size increase for treatment effects that are not clinically meaningful. Hence, the selection of $\text{CP}_{\text{min}}$ depends on the cost versus power trade-off, and case-by-case simulation is suggested. Assuming the ``starting small and asking for more" scenario, sample size decrease is not considered in our design.

\subsection{SSR with multiple contrasts}
The power of contrast test is dependent on the true D-R profile. A single contrast test may not be sufficiently powerful for all possible D-R scenarios, since the true D-R curve is hard to precisely estimate with limited pre-clinical and Phase I data. Consequently, there arises benefit to conduct multiple contrast tests together to limit the loss in power due to suboptimal selection of contrasts. One classic example is the Dunnett's test, which compares $k-1$ treatment arms with a single control. \citet{wang_2010} studied the performance of Dunnett's test in SSR designs. Another example is the MCPMod \citep{pinheiro_2006}, which combines optimal contrasts of multiple parametric D-R model candidates for PoC. In this part, we  follow the general methodology introduced by \citet{miller_2010} for extending the unblinded SSR design to trend test with multiple contrasts. 

Suppose $m$ contrasts are tested. Let $\boldsymbol{C}$ denote the $m\times k$ matrix of contrast coefficients, of which each row represents one contrast. Denote the sample mean vector by $\overline{\boldsymbol{Y}}$, which is $k$-dimensional. The covariance matrix of $\overline{\boldsymbol{Y}}$ is diagonal, namely $\boldsymbol{\Sigma}=\frac{\sigma^2}{N}\cdot\text{diag}(1/\phi_1,\ldots,1/\phi_k)$. Define matrix $\boldsymbol{W}$ as an $m\times m$ diagonal matrix, in which for the $j$th contrast test, the diagonal element $\boldsymbol{W}_{jj}=\sum_{i=1}^k \boldsymbol{C}_{ji}^2/\phi_i$. Combining all $m$ contrast tests, the test statistics can be computed together through
\begin{equation}
\label{correlation}
\boldsymbol{T}=\frac{\sqrt{N}}{\sigma}\cdot\boldsymbol{W}^{-1/2}\boldsymbol{C}\overline{\boldsymbol{Y}}\sim\mathcal{N}\left(\frac{\sqrt{N}}{\sigma}\cdot\boldsymbol{W}^{-1/2}\boldsymbol{C}\boldsymbol{\mu},\; \frac{N}{\sigma^2}\cdot\boldsymbol{W}^{-1/2}\boldsymbol{C}\boldsymbol{\Sigma}\boldsymbol{C}^\top\boldsymbol{W}^{-1/2}\right), 
\end{equation}
which follows multivariate normal distribution. It is noteworthy that $N$ in the covariance matrix can be cancelled out, so the correlation structure indicated by $\boldsymbol{\Theta}$ is invariant to sample size \citep{bornkamp_2009}. The rejection region can be established by comparing the maximum of the $m$ test statistics with the multiplicity adjusted critical value, controlling the familywise type I error at $\alpha$. 

As the multiple contrast test is extended to a two-stage scenario, we still adopt the same set of contrasts in stage 2. Without SSR, we conduct the final analysis by means of weighted combination statistics
\begin{equation*}
\boldsymbol{T}^*=r^{1/2}\cdot\boldsymbol{T}_1+(1-r)^{1/2}\cdot\boldsymbol{T}_2, 
\end{equation*}
where vectors $\boldsymbol{T}_1$ and $\boldsymbol{T}_2$ are independent. The critical value $u_\alpha$ is located to control familywise type I error, i.e.
\begin{equation*}
\alpha=\text{Pr}_{H_0}(\max(\boldsymbol{T})>u_\alpha). 
\end{equation*}
Under $H_0$, $\boldsymbol{T}$ is a zero mean $m$-variate normal vector with the same correlation structure $\boldsymbol{\Theta}$ as in formula \eqref{correlation}. All multivariate normal probabilities and quantiles in this paper are computed by the R package $\mathtt{mvtnorm}$ \citep{genz_1992, bornkamp_2018}, which numerically approximates multi-dimensional Gaussian CDF. 

When data dependent sample size modification is applied during the interim analysis, refine the stage 2 test statistics $\widetilde{\boldsymbol{T}}_2$ using the modified stage 2 sample size $\widetilde{N}_2$, 
\begin{equation*}
\widetilde{\boldsymbol{T}}_2=\frac{\sqrt{\widetilde{N}_2}}{\sigma}\cdot\boldsymbol{W}^{-1/2}\boldsymbol{C}\overline{\boldsymbol{Y}}^{(2)}. 
\end{equation*}
Likewise, the vector of weighted combination statistics is defined as
\begin{equation*}
\widetilde{\boldsymbol{T}}^*=r^{1/2}\cdot\boldsymbol{T}_1+(1-r)^{1/2}\cdot\widetilde{\boldsymbol{T}}_2.
\end{equation*}
Due to the fact that alteration of stage 2 sample size only does not change the correlation structure of $\widetilde{\boldsymbol{T}}_2$ from $\boldsymbol{T}_2$, the familywise type I error is strictly protected under the rejection region $\max(\widetilde{\boldsymbol{T}})>u_\alpha$. From the perspective of the conditional error approach \citep{muller_2004}, regardless of the observed stage 1 outcome, the conditional familywise type I error of the adaptive test always matches the conditional error of the non-adaptive test. 

Next, we turn to the computation of CP for stage 2 sample size re-assessment in multiple testing situation. The treatment effect size with respect to the contrast matrix is denoted by $\boldsymbol{\delta}=\boldsymbol{C}\boldsymbol{\mu}$. Subsequently, the CP is formulated by the following formula
\begin{equation*}
\text{CP}(n_2,\boldsymbol{\delta}\mid\boldsymbol{Y}^{(1)})=1-\Psi_{n_2,\boldsymbol{\delta}}\left(u_\alpha\right), 
\end{equation*}
where $\Psi_{n_2,\boldsymbol{\delta}}$ is the equicoordinate CDF (probability of all entries of a random vector less than or equal to a common value) of the distribution of $\widetilde{\boldsymbol{T}}\mid\boldsymbol{Y}^{(1)}$ under the assumed alternative hypothesis, which follows multivariate normal distribution with mean vector
\begin{equation*}
r^{1/2}\cdot \boldsymbol{T}_1+(1-r)^{1/2}\frac{\sqrt{n_2}}{\sigma}\cdot \boldsymbol{W}^{-1/2}\boldsymbol{\delta}
\end{equation*}
and covariance matrix $(1-r)\cdot\boldsymbol{\Theta}$ independent of $n_2$ and $\boldsymbol{\delta}$. Apparently, in the multiple contrast test scenario, CP is still monotonic in $n_2$ if $\boldsymbol{\delta}$ is fixed, and is strictly increasing in $n_2$ given $\boldsymbol{\delta}$ is entry-wise positive. Apply the same ``promising zone'' design as in Section~\ref{sec:ssr}. When the interim $\text{CP}(N_2)$ falls within the ``promising zone'', we solve the equation $\text{CP}(n_2)=1-\beta$ to specify the appropriate sample size $\widetilde{N}_2$ for the next stage. 

In comparison with the SSR design using Dunnett's test proposed by \citet{wang_2010}, our SSR with multiple contrasts is very similar in terms of test statistics and type I error control. However, the SSR rule in \citet{wang_2010} is not derived from a power equation, so it cannot achieve an exact required power. In our work, this ``promising zone'' design is utilized to add more flexibility, and the sample size increase is determined by power calculation.

\section{Bayesian approach for SSR}\label{sec:bayes}
\subsection{Bayesian predictive power}\label{sec:bayesian_test}
For a long time, there has been criticism on the use of CP that the uncertainty of the true treatment effect is not taken into account \citep{spiegelhalter_1986}. In fact, CP is highly dependent on the assumption associated with the true treatment effect. If we use the point estimate derived from stage 1 data to evaluate CP, the CP function may suffer from the high variability in early interim results \citep{glimm_2012}. On the other hand, if we simply apply the originally specified treatment effect, it does not take the existing evidence obtained from the interim analysis into account, even if there is significant discrepancy between observed data and assumption. Compared with frequentist power calculation, Bayesian PP allows a pooling of prior information and emerging evidence in order to anticipate the stage 2 outcome. As a result, the use of Bayesian PP can be helpful in balancing the historical knowledge and the information observed from stage 1. 

Assuming the same ANOVA model for response variable and specifying a prior distribution $\pi(\boldsymbol{\mu})$, we can obtain the Bayesian posterior distribution $p(\boldsymbol{\mu}\mid\boldsymbol{Y}^{(1)})$ as well as $p(\boldsymbol{\delta}\mid\boldsymbol{Y}^{(1)})$ given observed interim data. Though the treatment effect is modeled from Bayesian perspective, we still employ the frequentist test to conduct the final confirmatory analysis, which maintains the statistical rigor in type I error protection. Accordingly, PP is adopted to re-calculate the sample size, which is the probability of observing a success in the future based on existing data and historical knowledge. Since it is assumed that stage 2 data are generated from the posterior predictive distribution, the posterior predictive power (PP) can be simply interpreted as the chance of trial success if the trial continues \citep{dmitrienko_2006}. To be specific, the PP in our study can be expressed as
\begin{equation}
\label{pp}
\text{PP}(n_2\mid\boldsymbol{Y}^{(1)})=\mathbb{E}_{\delta\mid\boldsymbol{Y}^{(1)}}\left[\text{CP}(n_2,\delta\mid\boldsymbol{Y}^{(1)})\right]. 
\end{equation}
From the above formulation, PP is viewed as an extension of CP, which is to average the previous CP over different possibilities of treatment effect using its posterior distribution. Without much effort, the formula can be slightly modified to define PP for multiple trend tests. 

As pointed out by \citet{dallow_2011}, different from CP, the PP function is not necessarily monotonic as the sample size increases. For example, when the stage 1 result has already been significant, it is possible to observe a U-shaped ``PP vs. $n_2$'' curve. Also, PP does not always increase to 100\% as the final sample size increases. Therefore, {\color{red} when PP is selected for SSR instead of CP}, special attention has to be paid to take care of these differences. From historical experience in literature, the sample size increase required to reach a high target PP level, such as 90\%, under a targeted effect may be substantially greater than the sample size targeting CP at the same level. Therefore, if a high PP target is selected, there might be considerable price to pay for the extra power.

\subsection{Sample size re-estimation}
An analogy is drawn between the Bayesian SSR and the frequentist SSR criterion. The interim analysis is still implemented after a sample size of $N_1$. Similar to the CP-based design in frequentist statistics, we evaluate the posterior predictive power at $N_2$ to check whether trial success is likely following the original design. In case that the PP curve is not monotonic, we also look at the predictive power at {\color{red} $N_2=0$} to see whether stage 1 itself is powerful enough. We follow the same idea of ``promising zone" design associated with CP, and the re-estimation algorithm based on PP is summarized by the following three zones: 
\begin{enumerate}
\item Unfavorable: $\text{PP}(N_2)<\text{PP}_{\text{min}}$ and $\text{PP}(0)<\text{PP}_\text{min}$, no sample size increase is required. 
\item Promising: Otherwise, we select the minimum stage 2 sample size $\check{N}_2$ whose PP can achieve $1-\beta$, that is
\begin{equation*}
\check{N}_2=\min\left(\{n_2: \text{PP}(n_2\mid\boldsymbol{Y}^{(1)})\geq 1-\beta\}\right).
\end{equation*}
Taking the same practical limit into account, our final choice of of stage 2 sample size is $\widetilde{N}_2=\min(\check{N}_2, N_{\text{max}})$. If the PP can never achieve $1-\beta$ as $n_2$ increases, we let $\widetilde{N}_2=N_{\text{max}}$. 
\item Favorable: $\text{PP}(N_2)\geq 1-\beta$ or $\text{PP}(0)\geq 1-\beta$, the sample size is maintained at $N_2$. 
\end{enumerate}

\subsection{Bayesian computation}\label{sec:computation}
Motivated by Bayesian MCPMod introduced by \citet{fleischer2022bayesian}, we discuss methods of computing PP under different prior specifications for SSR in this part. In some cases when non-informative or conjugate priors are adopted, the corresponding posterior and predictive distributions can be derived analytically, which leads to efficient and accurate closed-form computation of PP. However, in general cases, Monte Carlo simulation or numerical approximation are applied to evaluate PP at some specific sample size. 

\subsubsection{Non-informative prior}\label{sec:noninfo}
First, we assume an improper non-informative prior $\pi(\boldsymbol{\mu})\propto 1$. Following the usual notations of ANOVA, the posterior distribution of $\boldsymbol{\mu}$ follows
\begin{equation*}
\mu_i\mid \boldsymbol{Y}^{(1)}\overset{\text{ind.}}{\sim} \mathcal{N}\left(\overline{Y}_{i\cdot}^{(1)}, \frac{\sigma^2}{N_1\phi_i}\right), \; i=1,\ldots,k. 
\end{equation*}
Then, from the normal-normal model, provided $\widetilde{N}_2=n_2$, the posterior predictive distribution of observed stage 2 treatment effect is given by
\begin{equation*}
\sum_{i=1}^k c_i\overline{Y}_{i\cdot}^{(2)}\mid\boldsymbol{Y}^{(1)}\sim\mathcal{N}\left(\sum_{i=1}^k c_i\overline{Y}_{i\cdot}^{(1)},\left(\frac{\sigma^2}{N_1}+\frac{\sigma^2}{n_2}\right)\sum_{i=1}^k c_i^2/\phi_i\right). 
\end{equation*}
Let us denote the posterior predictive mean of stage 2 treatment effect by $\delta^*$, which equals $\sum_{i=1}^k c_i\overline{Y}_{i\cdot}^{(1)}$ in this example. Then, the posterior predictive distribution of final test statistic $\widetilde{T}$ can be written as $\widetilde{T}\mid\boldsymbol{Y}^{(1)}\sim\mathcal{N}(a^*,b^*)$, where
\begin{align*}
a^*&=r^{1/2}\cdot T_1+(1-r)^{1/2}\cdot\frac{\delta^*\sqrt{n_2}}{\sigma\sqrt{\sum_{i=1}^k  c_i^2/\phi_i}}, \\
b^*&=(1-r)\cdot\left(1+\frac{n_2}{N_1}\right). 
\end{align*}
Based on the definition of Bayesian PP, we are able to derive $\text{PP}(n_2\mid\boldsymbol{Y}^{(1)})=1-\Phi\left[(z_\alpha-a^*)/\sqrt{b^*}\right]$.  Due to the extra variation brought by the Bayesian modeling of $\delta$, given the same observed stage 1 outcome, the value of $\text{PP}(n_2\mid\boldsymbol{Y}^{(1)})$ is closer to 0.5, compared with $\text{CP}(n_2,\widehat{\delta}\mid\boldsymbol{Y}^{(1)})$\citep{lan_2009}. 

Next, we generalize our result to multiple trend tests. From the same posterior distribution of $\boldsymbol{\mu}$, we are able to find the posterior predictive distribution of $\overline{\boldsymbol{Y}}^{(2)}$ follows a multivariate normal distribution, with mean vector $\overline{\boldsymbol{Y}}^{(1)}$ and covariance matrix 
\begin{equation*}
\boldsymbol{\Sigma}_2^*(n_2)=\text{diag}\left(\frac{\sigma^2}{N_1\phi_1}+\frac{\sigma^2}{n_2\phi_1}\;,\ldots,\;\frac{\sigma^2}{N_1\phi_k}+\frac{\sigma^2}{n_2\phi_k}\right).  
\end{equation*}
Accordingly, the posterior predictive distribution of the observed treatment effect is written as
\begin{equation*}
\boldsymbol{C}\overline{\boldsymbol{Y}}^{(2)}\mid\boldsymbol{Y}^{(1)}\sim\mathcal{N}\left(\boldsymbol{\delta}^*,\boldsymbol{C}\boldsymbol{\Sigma}_2^*(n_2)\boldsymbol{C}^{\top}\right), 
\end{equation*}
where $\boldsymbol{\delta}^*=\boldsymbol{C}\overline{\boldsymbol{Y}}^{(1)}$. Thereby, the posterior predictive distribution of $\widetilde{\boldsymbol{T}}$ is also multivariate Gaussian with mean $\boldsymbol{a}^*$ and covariance $\boldsymbol{B}^*$, where
\begin{align*}
\boldsymbol{a}^*&=r^{1/2}\cdot\boldsymbol{T}_1+(1-r)^{1/2}\cdot\frac{n_2}{\sigma}\cdot[\widetilde{\boldsymbol{W}}_2(n_2)]^{-1/2}\boldsymbol{\delta}^*, \\
\boldsymbol{B}^*&=(1-r)\cdot\left(1+\frac{n_2}{N_1}\right)\cdot\boldsymbol{\Theta}. 
\end{align*}
Thus, for multiple contrast tests, PP under non-informative prior can also be calculated through the CDF of multivariate normal distribution. Compared with CP using point estimate as treatment effect, the only difference appears at the additional factor in the covariance matrix. 

\subsubsection{Conjugate prior}
Next, we assume independent conjugate normal priors, i.e. $\mu_i\overset{\text{ind.}}{\sim}\mathcal{N}(\mu_{0i},\tau_{0i}^{-1})$ for $i=1,\ldots,k$. We use $\tau=\sigma^{-2}$ to denote the normal precision of our model. The posterior distribution of $\boldsymbol{\mu}$ is
\begin{equation*}
\mu_i\mid \boldsymbol{Y}^{(1)}\overset{\text{ind.}}{\sim} \mathcal{N}\left(\frac{\tau_{0i}\mu_{0i}+\tau N_1\phi_i\overline{Y}_{i\cdot}^{(1)}}{\tau_{0i}+\tau N_1\phi_i}, \left(\tau_{0i}+\tau N_1\phi_i\right)^{-1}\right), \;i=1,\ldots,k. 
\end{equation*}
The posterior predictive distribution of stage 2 observed treatment effect becomes
\begin{align*}
\sum_{i=1}^k c_i\overline{Y}_{i\cdot}^{(2)}\mid\boldsymbol{Y}^{(1)}&\sim\mathcal{N}\left(\sum_{i=1}^k c_i\frac{\tau_{0i}\mu_{0i}+\tau N_1\phi_i\overline{Y}_{i\cdot}^{(1)}}{\tau_{0i}+\tau N_1\phi_i},\sum_{i=1}^k c_i^2\left[\left(\tau_{0i}+\tau N_1\phi_i\right)^{-1}+\frac{\sigma^2}{n_2\phi_i}\right]\right),  
\end{align*}
Still use $\delta^*$ to denote the new posterior predictive mean of stage 2 treatment effect, and the posterior predictive power $\text{PP}(n_2\mid\boldsymbol{Y}^{(1)})=1-\Phi\left[(z_\alpha-a^*)/\sqrt{b^*}\right]$ can be reformulated with new
\begin{align*}
a^*&=r^{1/2}\cdot T_1+(1-r)^{1/2}\cdot\frac{\delta^*\sqrt{n_2}}{\sigma\sqrt{\sum_{i=1}^k  c_i^2/\phi_i}}, \\
b^*&=(1-r)\cdot\left(1+\frac{n_2\sum_{i=1}^k c_i^2\left(\tau_{0i}+\tau N_1\phi_i\right)^{-1}}{\sigma^2\sum_{i=1}^k c_i^2 \phi_i^{-1}}\right). 
\end{align*}
We can find that as the prior precision $\tau_{0i}$ grows to infinity for all $i=1,\ldots,k$, the PP coincides with CP whose $\delta$ is determined by the prior mean. On the other hand, as the prior precision $\tau_{0i}$ approaches 0 for all arms, the result is exactly the same as the one using non-informative prior. 

While in the multiple testing situation, using the same posterior distribution of $\boldsymbol{\mu}$, we also modify the posterior predictive distribution $\boldsymbol{C}\overline{\boldsymbol{Y}}^{(2)}\mid\boldsymbol{Y}^{(1)}$ with newly defined $\boldsymbol{\delta}^*$ and 
\begin{equation*}
\boldsymbol{\Sigma}_2^*(n_2)=\text{diag}\left(\left(\tau_{01}+\tau N_1\phi_1\right)^{-1}+\frac{\sigma^2}{n_2\phi_1}\;,\ldots,\;\left(\tau_{0k}+\tau N_1\phi_k\right)^{-1}+\frac{\sigma^2}{n_2\phi_k}\right). 
\end{equation*}
As a consequence, under the conjugate prior, the PP of the two-stage adaptive testing procedure is still able to expressed as a normal probability. 

\subsubsection{Generial prior distributions}
In general, the prior distribution $\pi(\boldsymbol{\mu})$ can be any distribution, either proper or improper that leads to a proper posterior. Also, it is not necessary for $\mu_1,\ldots,\mu_k$ to be independent in the specification of prior distribution.  In cases where a closed form expression of the predictive distribution is unavailable, Markov Chain Monte-Carlo may be used to sample from the posterior distribution. Each sample from the posterior distribution is used to calculate the CP. The predictive power is estimated as the average CP over the samples from the posterior distribution \citep{robert_2013}. To avoid repeated sampling from the posterior distribution in the adaptive sample size calculation, the conditional power should be calculated for each sample using all admissible stage 2 sample sizes.

Laplace approximation may be useful to avoid sampling from the posterior distribution in large scaled simulation problems. The resulting approximation to the posterior is again a multivariate normal distribution, allowing application of formulae as described in the previous section.

\section{Simulation studies} \label{sec:simu}
\subsection{Simulation settings}
\begin{figure}[h]
\centering
\includegraphics[width=0.8\textwidth]{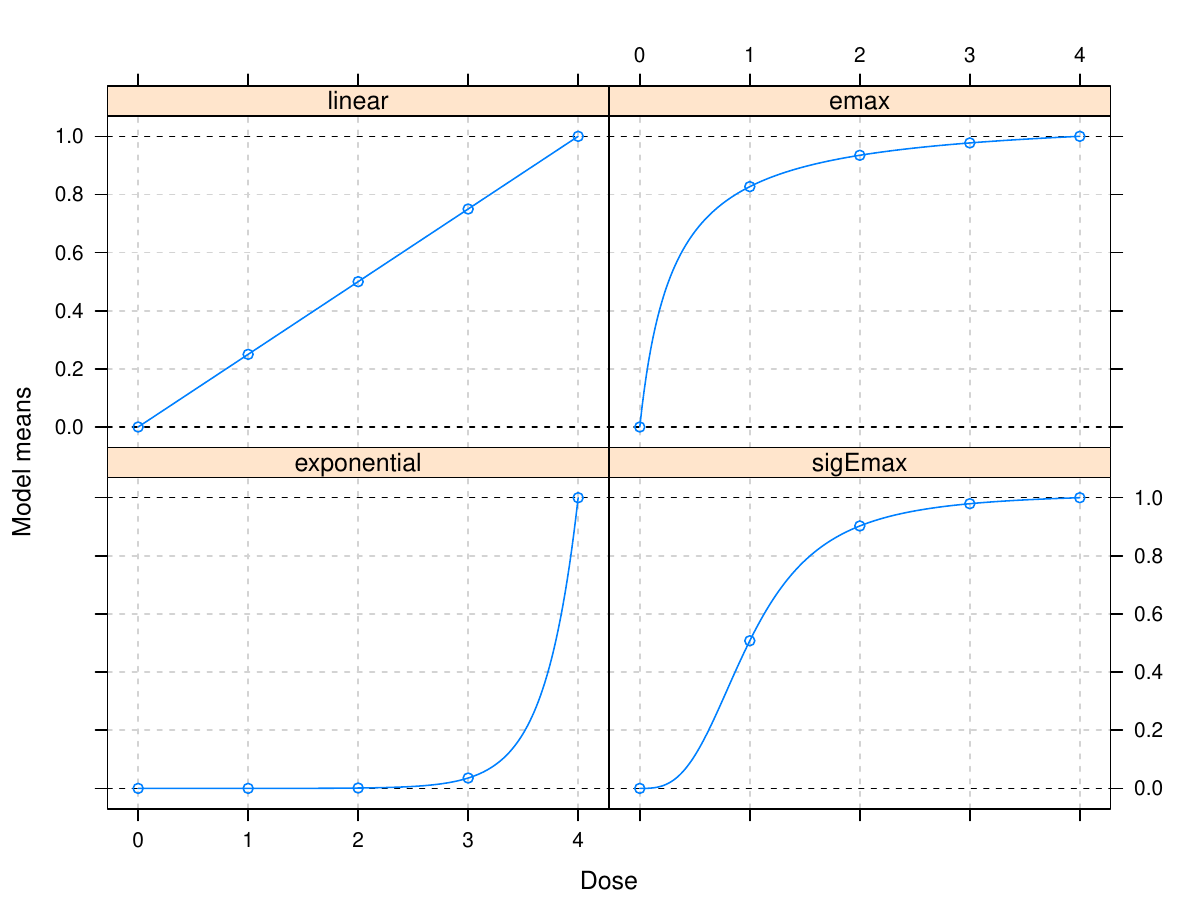}
\caption{The candidate parametric dose-response models used for multiple comparison. }\label{fig:mcp}
\end{figure}

 Let us assume that in a dose ranging trial with 5 equally allocated treatment groups, there are one placebo arm and 4 active doses, namely $0,1,\ldots,4$. The D-R relationship is assumed linear, and the within-group standard deviation is $\sigma=2$. At the trial design stage, we also suspect other potential D-R shapes, so we carry out a two-stage multiple contrast test to ensure the stability of power under different alternatives. The four candidate shapes are specified as the following standardized parametric models: Linear, Emax ($\text{ED50}=0.3$), Exponential ($\delta=0.3$), and Sigmoid Emax ($\text{ED50}=1$, $h=3$). These candidates D-R models are displayed in Figure~\ref{fig:mcp}. Using equal allocation proportion, the optimal contrast matrix is
\begin{equation*}
\boldsymbol{C}=
\begin{bmatrix}
-0.632 & -0.316 & 0 & 0.316 & 0.632 \\
-0.883 & 0.093 & 0.221 & 0.271 & 0.298 \\
-0.234 & -0.234 & -0.232 & -0.194 & 0.894 \\
-0.792 & -0.199 & 0.262 & 0.352 & 0.376
\end{bmatrix}. 
\end{equation*}
The original design is formed by an optimistic scenario, where the mean vector is $(0,0.25,0.5,0.75,1)^\top$. With one-sided familywise type I error rate controlled at $10\%$, we are able to calculate that $N=34\times 5=170$ can achieve a power of $1-\beta=80\%$. In our simulation study, we also evaluate how the time of conducting interim analysis impacts the result, so we mainly consider the following two time points for interim analysis: 
\begin{itemize}
\item Early (around 40\%): $N_1=70$, $N_2=100$, $\boldsymbol{\phi}=(0.2,\ldots,0.2)^\top$; 
\item Late (around 70\%): $N_1=120$, $N_2=50$, $\boldsymbol{\phi}=(0.2,\ldots,0.2)^\top$. 
\end{itemize}
To better focus our study on the outcome of adaptive sample size, we do not consider contrast or allocation modification. We assume the worst acceptable efficacy scenario is $\boldsymbol{\mu}=(0,0.2,0.4,0.6,0.8)^\top$. Under this model that is less efficacious than the optimistic one, the multiple contrast test power of the initial design is only 66\%, and a sample size $N=265$ is now required to achieve 80\% of power in this case. Let the maximum sample size for stage 2 be $N_{\text{max}}=N_2+95$. 

To evaluate the performance of SSR given the drug is not as efficacious as our initial anticipation, we first let the true mean vector used for data generation be $\boldsymbol{\mu}=(0,0.2,0.4,0.6,0.8)^\top$. Then, unblinded SSR is expected to outperform the fixed design. For all types of designs, the ``promising zone'' lower bounds are specified as $\text{CP}_{\text{min}}=\text{PP}_{\text{min}}=30\%$. Both frequentist CP and Bayesian PP are tested for our unblinded SSR design, and the six different power calculation settings are summarized as follows: 
\begin{itemize}
\item FQ1: CP with $\boldsymbol{\delta}=\boldsymbol{C}\overline{\boldsymbol{Y}}^{(1)}$;
\item FQ2: CP with $\boldsymbol{\delta}=\boldsymbol{C}(0,0.25,0.5,0.75,1)^\top$;
\item FQ3: CP with $\boldsymbol{\delta}=\boldsymbol{C}(0,0.2,0.4,0.6,0.8)^\top$;
\item BY1: PP under non-informative prior;
\item BY2: PP under independent conjugate prior, with $\boldsymbol{\mu}_0=(0,0.25,0.5,0.75,1)^\top$ and $\boldsymbol{\tau}_{0}=5\cdot\boldsymbol{1}$;
\item BY3: PP under independent conjugate prior, with $\boldsymbol{\mu}_0=(0,0.2,0.4,0.6,0.8)^\top$ and $\boldsymbol{\tau}_{0}=5\cdot\boldsymbol{1}$. 
\end{itemize}
For BY2 and BY3, the prior distributions have effective sample size of 100 across all treatment groups, which are considered moderately informative at interim. Among these methods, FQ1 and BY1 are completely objective and driven by observed data. Methods FQ2 and BY2 utilize prior information to different extents, but the prior knowledge is based on the overly optimistic anticipation. Methods FQ3 and BY3 are formulated from the true model, which are used for analyzing sensitivity of these methods when altering prior specification. 

In this simulation study, we intend to evaluate the general performance of these different methods from the following aspects. The first one is the distribution of the three ``zones'' at the interim look. To have more insight into the SSR decision rule, we also explore the detailed distributions of $\text{CP}(N_2)$ or $\text{PP}(N_2)$. Another perspective is the trade-off between overall power gain and sample size increase, so that we are able to better balance the benefit and cost of SSR. Thus, we compare the following two evaluation metrics, average total sample size (Mean SS.) as well as the average sample size increase given that the stage 1 result is within the ``promising zone'' (Mean Incr.) among all methods. Additionally, it is valuable to assess the robustness of CP/PP approaches at different SSR timings, given that the historical knowledge may be inaccurate. Smaller differences in operating characteristics among overly optimistic and true prior information are preferred.

\subsection{Results}
\begin{figure}[h]
\centering
\includegraphics[width=0.98\linewidth]{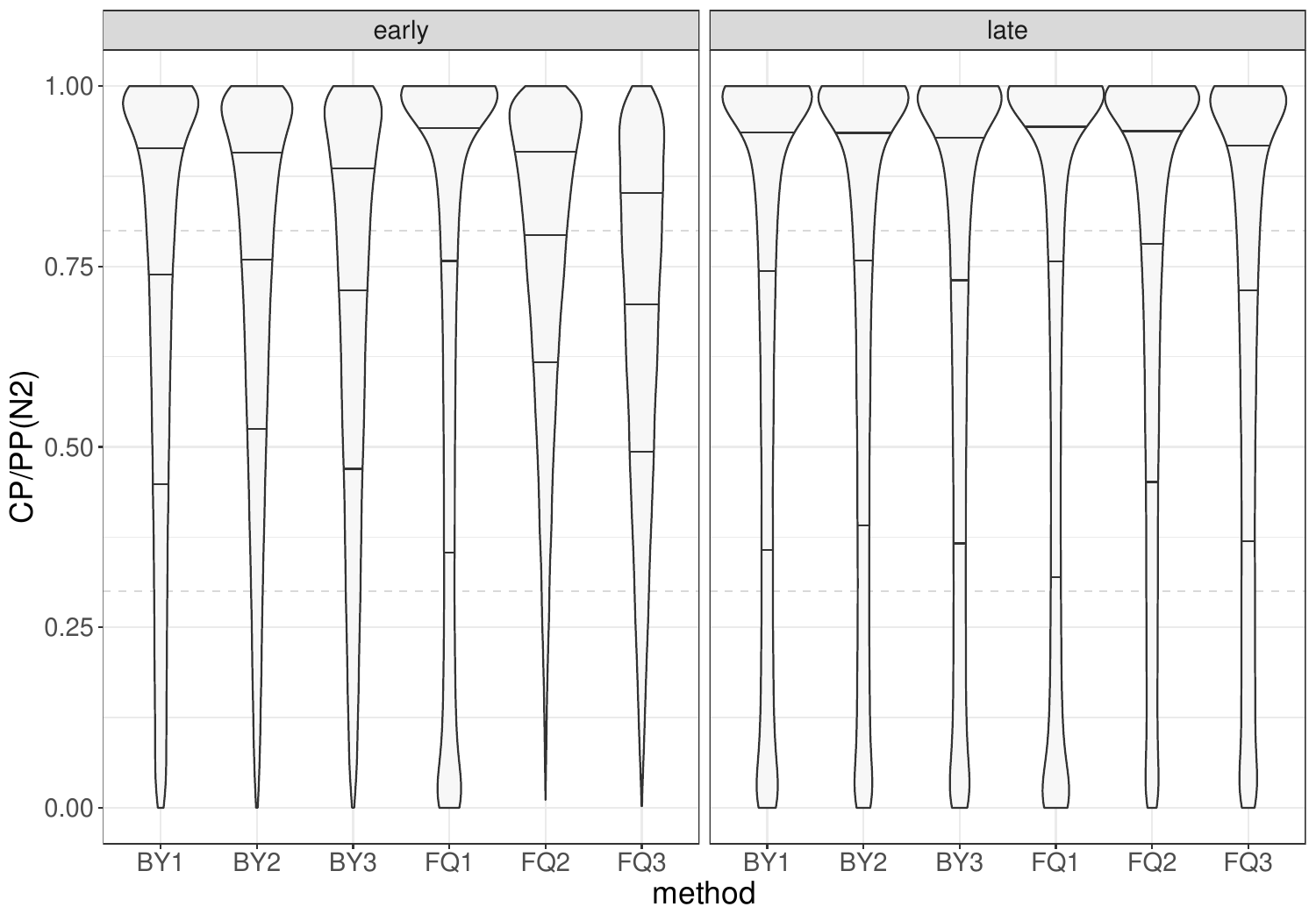}
\caption{Distributions of $\text{CP}(N_2)$ or $\text{PP}(N_2)$ given the true effect is below the optimistic scenario. The 25th, 50th and 75th percentiles of each distribution are indicated by solid lines inside the boxes. }\label{fig:cppp}
\end{figure}

\begin{figure}[h]
\centering
\includegraphics[width=0.98\linewidth]{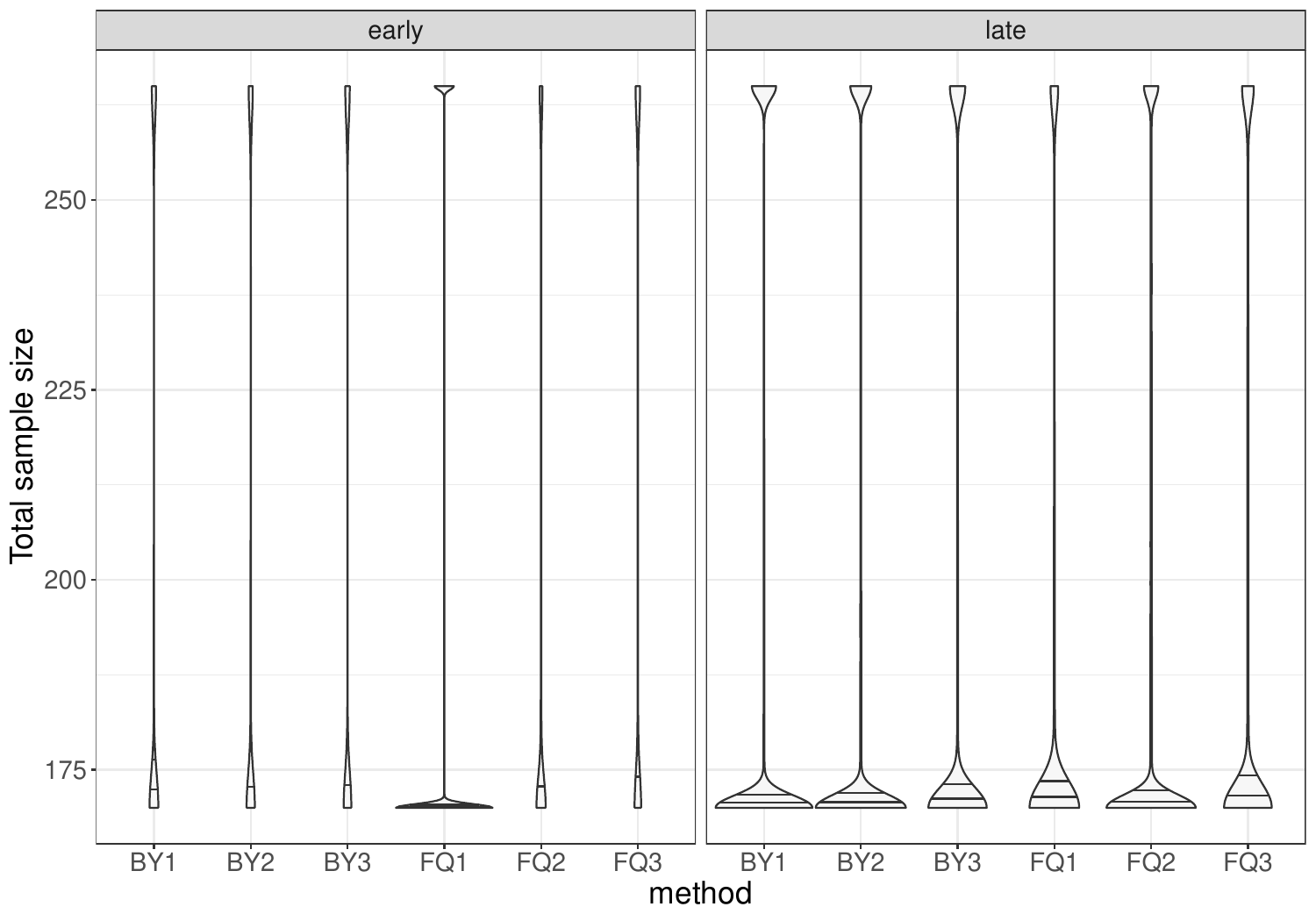}
\caption{Distributions of total sample size given the true effect is below the optimistic scenario. The 25th, 50th and 75th percentiles of each distribution are indicated by solid lines inside the boxes. }\label{fig:ss}
\end{figure}

In each scenario, we simulate 50,000 trials and the result summary is exhibited in Table~\ref{tab:mcp}. The distributions of interim $\text{CP}(N_2)$ or $\text{PP}(N_2)$ are demonstrated by violin boxplots in Figure~\ref{fig:cppp},  and the distributions of total sample size are in Figure~\ref{fig:ss}. 
\begin{table}[h]
\centering
\caption{Simulation result of unblinded SSR on multiple contrast test using 50,000 replicates, given the true efficacy is below the planned effect. }\label{tab:mcp}
\begin{tabular}{llcccccccc}
\toprule
                       &     & \multicolumn{3}{c}{\% Zone} & \multicolumn{2}{c}{CP/PP$(N_2)$} & \multicolumn{1}{l}{} & \multicolumn{1}{l}{} &            \\
              &     & Unfavor.  & Prom.  & Favor. & Mean            & SD             & Power                & Mean SS.             & Mean Incr. \\ \midrule
\multirow{6}{*}{Early} & FQ1 &  21  &  26  &  54  &  0.68  &  0.35       &  0.70  &   188   &   72   \\
                       & FQ2 &  4  &  45  &  51  &  0.75  &  0.21  &  0.72  &   197   &  61    \\
                       & FQ3 &  9  &  55  &  35  &  0.66  &  0.23   &  0.75  &   210   &  73    \\ [0.7em]
                       & BY1 &  15  &  38  &  47  &  0.68  &  0.29  &  0.72  &   202   &  82    \\
                       & BY2 &  9  &  43  &  48  &  0.71  &  0.25  &  0.73  &  203  &  77    \\
                       & BY3 &  12  &  46  &  42  &  0.68  &  0.26  &  0.74  &  206  &  79    \\ \midrule
\multirow{6}{*}{Late}  & FQ1 &  23  &  24  &  53  &  0.67  &  0.36       &  0.70  &   186   &   67   \\
                       & FQ2 &  16  &  31  &  53  &  0.71  &  0.31  &  0.71  &   189   &  61    \\
                       & FQ3 &  20  &  33  &  47  &  0.66  &  0.33  &  0.72  &  193  &  68    \\  [0.7em]
                       & BY1 &  21  &  28  &  51  &  0.67  &  0.34  &  0.71  &   191   &  75   \\
                       & BY2 &  19  &  29  &  52  &  0.68  &  0.33  &  0.72  &  191  &  71    \\
                       & BY3 &  20  &  30  &  49  &  0.67  &  0.34  &  0.71  &  192  &  73    \\ \bottomrule
\end{tabular} \\
\footnotesize{* Under the true model, the power of a fixed design is 0.66, 0.69, 0.73 for $N=170, 190, 210$, respectively.} 
\end{table}

Apparently, the simulation result uncovers the benefit of conducting unblinded SSR. Given that the true treatment effect is actually below expectation, all eight methods could significantly increase the probability of trial success at some cost of extra average sample size compared with the original fixed design with only 66\% power. For cases falling within the ``promising zone'', the power increases by a large margin. Compared with FQ2 and FQ3, the distribution of $\text{CP}(N_2)$ for method FQ1 has much greater variation. For FQ1, the source of variation in CP is not only the stage 1 test statistic, but also includes the estimation of $\delta$ depending on stage 1 outcome. The CP evaluated at the observed effect (FQ1) suffers from the problem that the observed effect is utilized twice: as the conditioning variable and as the variable for calculating the final test statistic \citep{glimm_2012}. Due to the possibility of poor stage 1 result, without utilizing historical information, FQ1 has much higher percentage of ``unfavorable zone'' and lower power compared with the other two frequentist methods. Since FQ2 is based on some optimistic prior knowledge, the percentage of ``unfavorable zone'' is less than that of FQ3, and the percentage of ``favorable zone'' is much higher. Taking advantage of correct prior information, FQ3 has the highest ``promising zone'' percentage as well as overall power among all CP-based methods. Overall, incorporating reliable historical knowledge into power calculation is helpful in improving precision of interim decision and SSR efficiency. 

Comparing the results between early and late interim analyses, we can observe that for FQ2 and FQ3, late interim analysis has greater variation in $\text{CP}(N_2)$, because longer stage 1 increases the weight of uncertain stage 1 test statistic. With more data gathered at interim analysis, the interim decisions tend to be more conclusive, so the percentages of ``promising zone'' drop substantially. For all CP-based methods, these percentages vary from 24\%-33\%, and they also have more homogeneous performance in power and sample size compared with early SSR. Comparing FQ2 and FQ3 as well as BY2 and BY3, we can observe that with more objective information accumulated, the adaptive design becomes more resistant to inaccurate historical knowledge. Other than robustness, late interim analysis also leads to better power efficiency in general. For example, both early and late interim analyses achieve similar level of power for FQ1 and FQ2, but late interim analysis usually requires smaller sample size on average. 

Then, we make a parallel comparison between all Bayesian power methods (BY1, BY2, BY3) and all frequentist power methods (FQ1, FQ2, FQ3). The overall performance of late FQ1 and BY1 are comparable, but BY1 has slightly greater ``promising zone'' percentage, higher power and mean sample size. For the other methods with informative prior, $\text{PP}(N_2)$ has greater variation than the corresponding frequentist methods, but it is less variable than that of non-informative prior. The overall trends among the three Bayesian methods are similar to frequentist approaches, but the heterogeneity among operating characteristics under different prior information is much reduced. Especially, given moderately informative prior, powers of Bayesian methods are more robust compared with the results of FQ2 and FQ3, when there exists discrepancy between prior and truth. As for the power sample size trade-off, because it is more difficult for PP to achieve a high number, all Bayesian PP methods are more conservative and increase more sample size than frequentist methods with less variation in CP calculation. 

\begin{figure}[h]
\centering
\includegraphics[width=0.98\linewidth]{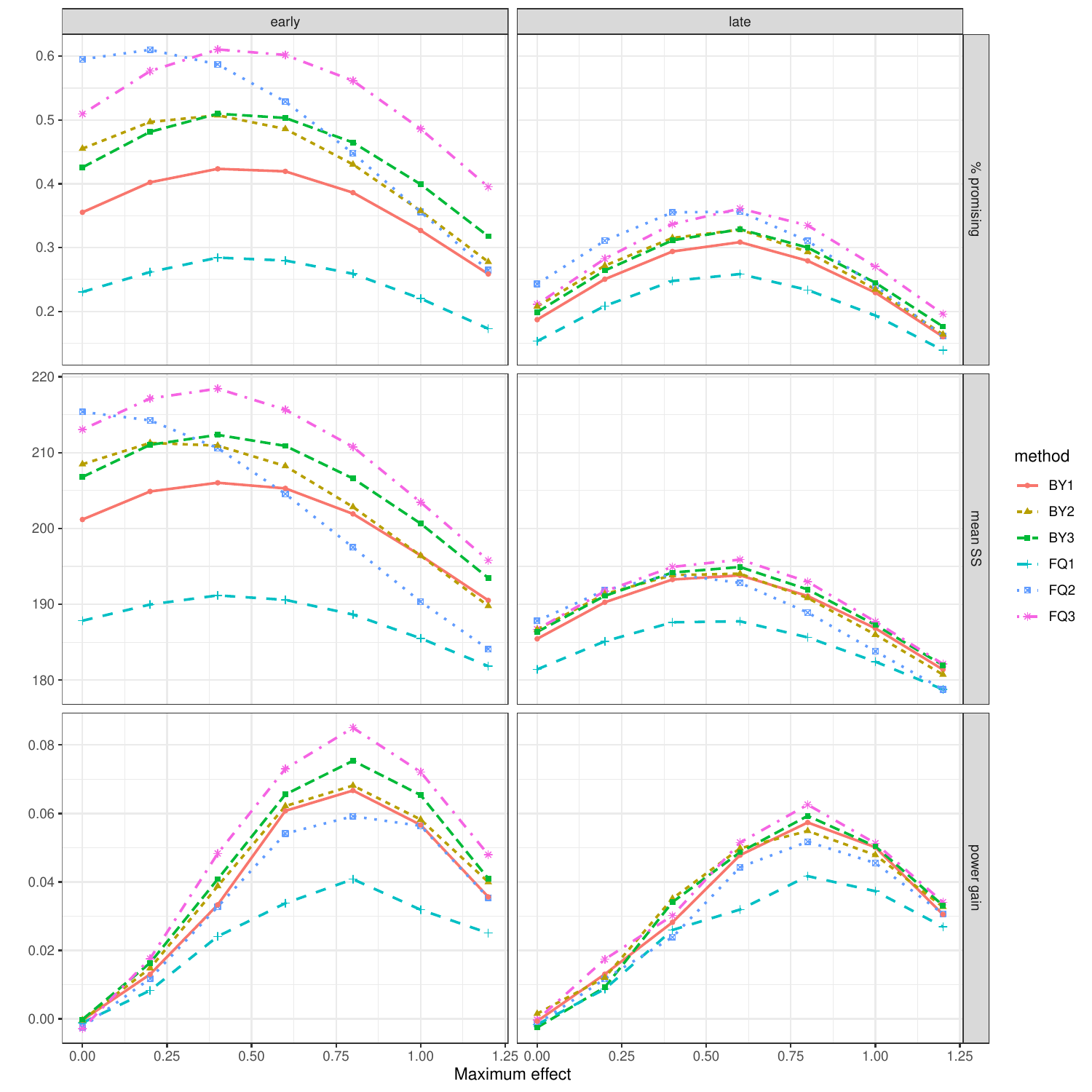}
\caption{Simulation result of unblinded SSR on multiple contrast test under different maximum effect assuming linear shape. Power gain equals the difference between the power of SSR and the power of fixed design with $N=170$. } \label{fig:metric}
\end{figure}

The property of our SSR design under the null, correct and super-efficacious scenarios is also of our interest, in which sample size increase is not necessary. Thus, we test the six power calculation settings under $\boldsymbol{\mu}=(0,0,0,0,0)^\top$, $(0,0.2,0.4,0.6,0.8,1)^\top$ as well as $\boldsymbol{\mu}=(0,0.3,0.6,0.9,1.2)^\top$. The percentage of rejection of null hypothesis, percentage of ``promising zone'', average total sample size and average sample size increase are summarized in Table~\ref{tab:mcp2}. The result highlights the necessity of late interim analysis. For all types of power calculation methods, conducting interim late substantially reduces the probability of unneeded sample size increase. Compared with the result in Table~\ref{tab:mcp}, the ``promising zone'' percentages and mean sample sizes are smaller in Table~\ref{tab:mcp2} for late SSR. Under the null setting, Bayesian methods BY2 and BY3 show better stability than FQ2 and FQ3, provided that the historical knowledge is unreliable and has a great gap from the truth. If we are skeptical about the prior information, Bayesian PP better balances the utilization of historical knowledge and robustness. Assuming linear D-R shape, more supporting information under various maximum effect sizes (0, 0.2, 0.4 , $\ldots$, 1.2) for different methods can be found in Figure~\ref{fig:metric}, from which we have similar findings. FQ1 leads to least sample size increase in all scenarios; FQ2 and FQ3 have more unnecessary sample size increase given ineffective treatment; Bayesian methods are more robust across different maximum effect scenarios in comparison with FQ2 and FQ3. 
\begin{table}[h]
\centering
\caption{Simulation result of unblinded SSR on multiple contrast test using 50,000 replicates. The true efficacy is either under the null, correct or super-effective scenario. }\label{tab:mcp2}
\begin{tabular}{lllcccccc}
\toprule
                       &                        &            & FQ1 & FQ2 & FQ3 & BY1 & BY2 & BY3 \\ \midrule
\multirow{8}{*}{Null}  & \multirow{4}{*}{Early} & \% Rej. & 10 & 10 & 10 & 10 & 10 & 10 \\
& & \% Prom.   & 23  & 60  & 51  & 36  & 46  & 43  \\
                       &                        & Mean SS.   & 188 & 215 & 213 & 201 & 208 & 207 \\
                       &                        & Mean Incr. & 77  & 76  & 84  & 88  & 85  & 86  \\ [0.7em]
                       & \multirow{4}{*}{Late}  & \& Rej. & 10 & 10 & 10 & 10 & 10 & 10 \\
                       & & \% Prom.   & 16  & 24  & 21  & 19  & 21  & 20  \\
                       &                        & Mean SS.   & 182 & 188 & 187 & 186 & 187 & 186 \\
                       &                        & Mean Incr. & 75  & 73  & 80  & 83  & 80  & 82  \\ \midrule
\multirow{8}{*}{Correct}  & \multirow{4}{*}{Early} & \% Rej. & 83 & 86 & 87 & 86 & 86 & 87 \\
&  & \% Prom.   & 22  & 36  & 49  & 33  & 36  & 40  \\
                       &                        & Mean SS.   & 186 & 190 & 203 & 196 & 196 & 201 \\
                       &                        & Mean Incr. & 70  & 57  & 69  & 81  & 74  & 77  \\ [0.7em]
                       & \multirow{4}{*}{Late}  & \% Rej. & 84 & 85 & 85 & 85 & 85 & 85 \\ 
                       & & \% Prom.   & 19  & 24  & 27  & 23  & 23  & 24  \\
                       &                        & Mean SS.   & 182 & 184 & 188 & 188 & 186 & 187 \\
                       &                        & Mean Incr. & 64  & 58  & 66  & 73  & 68  & 71  \\ \midrule
\multirow{8}{*}{Super} & \multirow{4}{*}{Early} & \% Rej. & 93 & 94 & 95 & 94 & 94 & 94 \\ 
& & \% Prom.   & 17  & 26  & 39  & 26  & 27  & 32  \\
                       &                        & Mean SS.   & 182 & 184 & 196 & 191 & 189 & 194 \\
                       &                        & Mean Incr. & 69  & 54  & 66  & 79  & 71  & 74  \\ [0.7em]
                       & \multirow{4}{*}{Late} & \% Rej. & 93 & 93 & 93 & 93 & 93 & 93 \\
                       & & \% Prom.   & 14  & 16  & 19  & 16  & 17  & 18  \\
                       &                        & Mean SS.   & 179 & 179 & 182 & 182 & 181 & 182 \\
                       &                        & Mean Incr. & 63  & 55  & 63  & 71  & 65  & 68 \\
                       \bottomrule
\end{tabular}
\end{table}

\section{Discussion}\label{sec:discuss}
Unblinded SSR is a popular tool to address the problem of power uncertainty caused by imprecise treatment effect assumption. In this paper, we have developed and evaluated unblinded SSR for single and multiple contrast tests, using different frequentist and Bayesian power approaches under rigorous false positive rate protection. Given the simulation results, the following recommendations can be made. First, irrespective of PP or CP, late interim analysis may be more reliable for SSR. Next, the PP using non-informative prior and CP at observed effect perform similarly. Among them, CP is more variable, while PP may lead to larger sample size. In addition, available reliable prior information on the treatment effect is beneficial for sample size recalculation. When prior information is incorporated, the Bayesian approach is favorable due to its resistance to wrongly specified prior. Practically, the target power for Bayesian PP should not be too high to prevent the adaptive sample size increase from being inefficient. 

As a rough framework, our proposed method can be extended in many aspects. The usage of contrast test is not restricted to normal endpoints. By exploiting the asymptotic normality of ML and GLS estimates, contrast test can be applied to different types of endpoints, such as binary, time-to-event and more complex models like adding covariates \citep{pinheiro_2014}. Unblinded SSR is still able to address the power issue in these cases. Besides, we are able to incorporate adaptive alterations of contrast set and allocation proportion into our SSR design, which may be helpful in clinical trial safety, ethics or power optimality. Please see \citet{miller_2010} for detailed discussions and experiments on how to adapt the design and contrast vector for stage 2. For instance, we can further assume a parametric D-R model to compute the frequentist and Bayesian power. During the interim analysis, we make inference on model parameters (either in frequentist or Bayesian way) to re-estimate the effect size. In addition to borrowing more shape information to optimize the contrast, the biggest advantage is that we are able to add new dose levels in stage 2, which makes the adaptive design even more flexible. 

It is also noteworthy that operational bias for the interim analysis may be a major concern of regulators. So in practice, a fully deterministic sample size calculation should be avoided to reduce the extent of access to interim results. For instance, the amount of sample size increase for the ``promising zone" may be divided into rough intervals to lower the chance of back calculation. These concerns regarding operation bias should be addressed in the DMC charter.

\section*{Data Availability Statement}
The data that support the findings of this study are available from the corresponding author upon reasonable request.
\section*{Acknowledgement}
We would like to express our special gratitude to Dr. Qiqi Deng from Moderna for her valuable comments and discussions. We also thank the editor, associate editor, and the reviewers for carefully reading the manuscript and providing valuable comments and suggestions to improve the paper. Dr. Hu’s research was partially supported by US NSF grants BCS-2152822 and DMS-2210371.

\bibliography{reference}
\end{document}